\documentclass[aps,prd,unsortedaddress,superscriptaddress,showpacs,nofootinbib,twocolumn,10pt]{revtex4-2} 
\usepackage[utf8]{inputenc}
\usepackage{array}
\usepackage{multirow}
\usepackage{graphicx,color}
\usepackage{relsize}
\usepackage{slashed}
\usepackage[normalem]{ulem}
\usepackage{tabu}
\usepackage{rotating}
\usepackage{bigstrut}
\usepackage{makecell}
\usepackage[dvipsnames]{xcolor}
\usepackage[colorlinks=true, linkcolor=blue, citecolor=Green, urlcolor=blue]{hyperref}
\usepackage{amsmath}
\usepackage{amssymb,bm}
\usepackage{amsthm}
\usepackage{mathrsfs}
\usepackage{array}
\usepackage[all]{xy}
\usepackage{euscript}
\usepackage{enumerate}
\usepackage{mathtools}
\usepackage{soul}
\usepackage{orcidlink}

\allowdisplaybreaks

\graphicspath{{Figs/}}  

\newcommand{\X}{X(3872)}
\newcommand{\jp}{J/\psi}

\newcommand{\itp}{\affiliation{CAS Key Laboratory of 
Theoretical Physics, Institute of Theoretical Physics,\\
Chinese Academy of Sciences, Beijing 100190, China}}
\newcommand{\ucas}{\affiliation{School of Physical Sciences, 
University of Chinese Academy of Sciences, Beijing 100049, 
China}}
\newcommand{\peng}{\affiliation{Peng Huanwu Collaborative 
Center for Research and Education, Beihang University, Beijing 
100191, China}}

\newcommand{\bonn}{\affiliation{Helmholtz Institut f\"{u}r Strahlen- und Kernphysik and Bethe Center for Theoretical Physics,\\ 
Universit\"{a}t Bonn, D-53115 Bonn, Germany}}

\newcommand{\scnt}{\affiliation{Southern Center for Nuclear-Science Theory (SCNT), Institute of Modern Physics,\\ 
Chinese Academy of Sciences, Huizhou 516000, China}}

\newcommand{\fzj}{\affiliation{Institute for Advanced Simulation (IAS-4), Forschungszentrum J\"ulich, D-52425 J\"ulich, Germany}}
\newcommand{\Tbilisi}{\affiliation{Tbilisi State University, 0186 Tbilisi, Georgia}}

\newcommand{\CHEP}{\affiliation{School of Physics and Center of High Energy Physics, Peking University, Beijing 100871, China}}

\newcommand{\ufpi}{\affiliation{Departamento de F\'isica, Universidade Federal do Piau\'i, 64049-550 Teresina, Piau\'i, Brasil}}
\newcommand{\Xspi}{X(3872)\to J/\psi \pi^+\pi^-}
\newcommand{\Xlpi}{X(3872)\to J/\psi \pi^+\pi^0\pi^-}

\newcommand{\Om}{\Omega(s)}
\newcommand{\twopi}{\pi^+\pi^-}
\newcommand{\threepi}{\pi^+\pi^0\pi^-}

\begin{document}

\title{Dispersive analysis of the isospin breaking in the $\Xspi$ and $\Xlpi$ decays}

\author{Jorgivan Morais Dias\orcidlink{0000-0002-0354-4711}}
\email{jorgivan.mdias@itp.ac.cn}
\itp\ufpi

\author{Teng Ji\orcidlink{0000-0003-0366-1042}}
\email{teng@hiskp.uni-bonn.de}
\bonn

\author{Xiang-Kun Dong\orcidlink{0000-0001-6392-7143}}
 \email{xiangkun@hiskp.uni-bonn.de}
\bonn

\author{Feng-Kun~Guo\orcidlink{0000-0002-2919-2064}}
\email{fkguo@itp.ac.cn}
\itp\ucas\peng\scnt

\author{Christoph Hanhart\orcidlink{0000-0002-3509-2473}}\email{c.hanhart@fz-juelich.de}
\fzj

\author{Ulf-G. Mei{\ss}ner\orcidlink{0000-0003-1254-442X}}\email{meissner@hiskp.uni-bonn.de}
\bonn\fzj\Tbilisi

\author{Yu Zhang}
\email{2110020124@hhu.edu.cn}
\affiliation{College of Mechanics and Engineering Science, Hohai University, Nanjing 211100, China}
\itp

\author{Zhen-Hua Zhang\orcidlink{0000-0001-6072-5378}}
  \email{zhhzhang@pku.edu.cn}
\CHEP\itp

\begin{abstract}
We analyze the latest LHCb data on the $\pi^+\pi^-$ 
spectrum in the isospin-violating $\Xspi$ decay, based on dispersion theory to 
deal with the $\pi\pi$ final state interactions. 
Additionally, the isospin breaking effects 
are properly introduced, 
allowing for a reliable and accurate extraction of the 
ratio, $R_X$, between the $X(3872)$ couplings to the $J/\psi \rho$ 
and $J/\psi \omega$ channels from the data. We find very good 
agreement with the LHCb data for the whole range of the $\pi^+\pi^-$ 
invariant mass, and $R_X$ is determined to be {$0.26\pm 0.03$}. Using this value, we make predictions for the $\pi^+\pi^0\pi^-$ 
mass distribution in the $\Xlpi$ process, which is currently 
accessible by the BESIII Collaboration, and update a prediction for the pole positions of the isovector partner states of the $X(3872)$, $W_{c1}$, with $I(J^{PC})=1(1^{++})$.
\end{abstract}

\maketitle

\section{Introduction}
\label{Sec:intro}

The discovery of $X(3872)$, also known as $\chi_{c1}(3872)$, 
in 2003 by the Belle Collaboration~\cite{Belle:2003nnu}
in the $J/\psi\pi^+\pi^-$ invariant mass spectrum from 
$B$ meson decays, produced in $e^+e^-$ collisions, inaugurated a new era in 
hadron spectroscopy physics. Shortly after its discovery, 
the CDF~\cite{CDF:2003cab} and D$\O$~\cite{D0:2004zmu} Collaborations also confirmed its 
existence in $p\bar{p}$ collisions. Since then, 
many other experiments have investigated its properties 
in various processes~\cite{Belle:2006olv,BaBar:2010wfc,BESIII:2013fnz,LHCb:2015jfc,ATLAS:2016kwu,BESIII:2019esk,LHCb:2020xds,LHCb:2020fvo,CMS:2021znk,BESIII:2022bse,LHCb:2013kgk,Belle:2005lfc,BESIII:2019qvy},
making it the best studied 
hadronic structure among the new hadrons that, like the
$X(3872)$, behave differently from what would be 
expected if their quark content were consistent with the conventional constituent quark model (see Refs.~\cite{Hosaka:2016pey,Esposito:2016noz,Guo:2017jvc,Olsen:2017bmm,Karliner:2017qhf,Kalashnikova:2018vkv,Brambilla:2019esw,Meng:2022ozq, Liu:2024uxn,Chen:2024eaq} for recent reviews).

The latest Particle Data Group average values for the mass and width of the $X(3872)$ are $(3871.64 \pm 0.06)$~MeV and $(1.19\pm0.21)$~MeV, respectively~\cite{ParticleDataGroup:2024cfk}. 
However, one should notice that they were obtained from averaging values extracted using the Breit-Wigner (BW) parametrization~\cite{Belle:2011vlx,LHCb:2020fvo,LHCb:2020xds}, which is not appropriate when a resonance is located near the threshold of a channel that it strongly couples to in the $S$-wave. 
Using a generalized  Flatt\'e parameterization~\cite{Hanhart:2007yq}, which takes into account the thresholds properly, the LHCb Collaboration obtained the 
mass and the visible width 
defined by the full width at half maximum of the $X(3872)$ using a fit to the line shape in the $J/\psi\pi^+\pi^-$ final state from $b$-hadron decays as $3871.69_{-0.04-0.13}^{+0.00+0.05}$~MeV and $0.22^{\,+\,0.07\,+\,0.11}_{\,-\,0.06\,-\,0.13}$~MeV, respectively~\cite{LHCb:2020xds}.
The line shape emerged from a pole located on the second sheet displaced only by $0.06-i 0.13$ MeV
from the $D^0\bar D^{*\, 0}$ threshold.
 Recently, the BESIII Collaboration reported the mass parameter and imaginary part of its pole as $\left(3871.63 \pm 0.13_{-0.05}^{+0.06}\right)$~MeV and $(-0.19 \pm 0.08_{-0.19}^{+0.14})$~MeV, respectively, from the processes $e^{+} e^{-} \rightarrow \gamma X(3872)$, $X(3872) \rightarrow D^0 \bar{D}^0 \pi^0$ and $\pi^{+} \pi^{-} J / \psi$~\cite{BESIII:2023hml}.
One sees one intriguing 
characteristic of the $X(3872)$, that is, its mass 
coincides with the $D^0\bar{D}^{*0}$ threshold at $(3871.69 \pm 0.07)$~MeV~\cite{ParticleDataGroup:2024cfk}. 
In view of the tiny phase spaces, its branching fraction into the 
$D^0\bar{D}^{*0}$ channel as well as into $D^0\bar{D}^{0}\pi^0$ 
are remarkably large~\cite{Belle:2008fma,Li:2019kpj,Braaten:2019ags,BESIII:2020nbj,BESIII:2023hml}, indicating a strong coupling of the $X(3872)$ to the $D\bar D^*$.

As no charged partner of the  $X(3872)$ has been reported so far~\cite{BaBar:2004cah,Belle:2011vlx}, the $X(3872)$ is expected to be an isoscalar in the isospin symmetric limit. 
However, in the isospin breaking world, the  mass eigenstate is a mixture of different isospin eigenstates.
Measurements on isospin breaking processes are crucial to determine how large the admixture is.
For the $X(3872)$,
in this sense relevant measurements are provided by its branching fractions decaying into the modes $J/\psi \pi^+\pi^-\pi^0$ and $J/\psi\pi^+\pi^-$, ${\mathcal{B}[X(3872) \rightarrow J / \psi 3\pi]}/{\mathcal{B}\left[X(3872) \rightarrow J / \psi \pi^{+} \pi^{-}\right]}$,
\begin{align}
\begin{cases} 1.0\pm0.4\pm 0.3 & \mbox{Belle},\\
    0.7 \pm 0.3\, (1.7 \pm 1.3) & 
    \mbox{BaBar } B^{+}\, (B^0)\, \mbox{events},  \\   
    1.6_{-0.3}^{+0.4} \pm 0.2 & \mbox{BESIII}\, ,\end{cases}
\label{Eq:ratio_br}
\end{align}
reported by the Belle~\cite{Belle:2005lfc}, BaBar~\cite{BaBar:2010wfc}, 
and BESIII~\cite{BESIII:2019qvy} Collaborations.\footnote{The cuts on the $3\pi$ invariant mass are $m_{3\pi}>0.75$~GeV for Belle~\cite{Belle:2005lfc}, $m_{3\pi}\in [0.74, 0.7965]$~GeV for the $B^+$ events and $\in[0.74, 0.8055]$~GeV for the $B^0$ events for BaBar~\cite{BaBar:2010wfc}, and $m_{3\pi}\in (0.71, 0.81)$~GeV for BESIII~\cite{BESIII:2019qvy}, respectively. One also notices that the $3\pi$ distribution in the selected region of the BaBar measurement peaks at around 0.76~GeV and is significantly broader than the $\omega$ width.}
Given the positive $C$ parity of the $\X$~\cite{LHCb:2013kgk}, $C$-parity conservation and Bose-Einstein statistics imply that the $\pi^+\pi^-$ pair in the $\jp\pi^+\pi^-$ final state must be an isovector, coming mainly from the $\rho^0$ meson.
Accordingly the 3$\pi$ channel is expected to be saturated by the $\omega$ meson.
It is worthwhile to notice that a large part of the isospin breaking comes from the huge phase space difference between the $\X\to \jp\omega$ and $\X\to \jp\rho^0$~\cite{Suzuki:2005ha}.
Thus, the true measure of the isospin breaking effects at the dynamical level should be, instead of the ratio of branching fractions in Eq.~\eqref{Eq:ratio_br}, the ratio between 
the $X(3872)$ couplings to the $J/\psi\rho$ and 
$J/\psi\omega$ channels, that is~\cite{Suzuki:2005ha}
\begin{align}
    R_X \equiv \frac{g_{X\psi\rho}}{g_{X\psi\omega}}. 
    \label{eq:RX_def} 
\end{align}
For studies of the isospin breaking in the multiquark (either molecular or nonmolecular) configurations of the $X(3872)$, see Refs.~\cite{Tornqvist:2003na,Swanson:2003tb,Tornqvist:2004qy,Maiani:2004vq,Terasaki:2007uv,Gamermann:2009uq,Gamermann:2009fv,Hidalgo-Duque:2012rqv,Li:2012cs,Takeuchi:2014rsa,Albaladejo:2015dsa,Maiani:2020zhr,Maiani:2017kyi,Wu:2021udi,Meng:2021kmi,Wang:2023sii}. 

Obtaining the value of $R_X$ reliably and accurately is 
of utmost importance for understanding the mechanism behind 
the observable in Eq.~\eqref{Eq:ratio_br} and the very nature 
of the $X(3872)$. In particular, $R_X$ has been utilized as a crucial input to determine the isoscalar and isovector low-energy constants (LECs) of the $D\bar{D}^{\ast}$ interactions~\cite{Gamermann:2009uq,Hidalgo-Duque:2012rqv,Albaladejo:2015dsa,Ji:2022uie,Zhang:2024fxy}, which can be used to predict the pole positions of the heavy quark spin partners~\cite{Hidalgo-Duque:2012rqv,Albaladejo:2015dsa,Baru:2016iwj,Ji:2022uie} and the isovector partner $W_{c1}$~\cite{Zhang:2024fxy} of the $X(3872)$ in the hadronic molecular picture.

The ratio $R_X$ was first estimated to be 
$0.29\pm0.02$~\cite{Suzuki:2005ha} and $0.30\pm0.07$~\cite{Braaten:2005ai} in 2005 using the experimental value of ${\mathcal{B}[X(3872) \rightarrow J / \psi \pi^+\pi^0\pi^-]}/{\mathcal{B}\left[X(3872) \rightarrow J / \psi \pi^{+} \pi^{-}\right]}$
from Belle~\cite{Belle:2005lfc}, where the two processes are mediated by the $\rho$ and $\omega$ resonances using the BW parameterization.
Such parameterization for the $\rho$ meson is precarious as the broad bump in the line shapes from the $\rho$ resonance cannot be well described by the BW function~\cite{Daub:2015xja}. In addition, $\rho$-$\omega$ mixing was shown to have a significant impact on the two-pion 
channel~\cite{Hanhart:2011tn}. This led to an improved value of the  pertinent ratio,
$R_X~=~0.26^{+0.08}_{-0.05}$,
 by fitting the data from Belle~\cite{Belle:2011vlx} and Babar~\cite{BaBar:2010wfc} 
on the invariant mass distributions of $\pi^+\pi^-$ 
and $\pi^+\pi^0\pi^-$ in the $\Xspi$ and $\Xlpi$ 
decays, respectively. 
The recent LHCb experiment~\cite{LHCb:2022jez} updated the $\pi^+\pi^-$ invariant mass distribution in $X(3872)\to J/\psi\pi^+\pi^-$ and estimated $R_X$ to be $0.29\pm0.04$ utilizing a similar strategy as in Ref.~\cite{Hanhart:2011tn}, but it set the $X(3872)$ mass to be 4~GeV, much larger than the Flatt\'e result~\cite{LHCb:2020xds}, to extend the phase space. 
In Ref.~\cite{Wang:2022vjm}, the updated LHCb data~\cite{LHCb:2022jez} for $\Xspi$ as well as previous BaBar data~\cite{BaBar:2010wfc} for $\Xlpi$ were analyzed simultaneously, where the $\omega$ meson contribution via 
$\omega \to \pi^+\pi^-$ was taken into account through a complex-valued 
effective coupling instead of $\rho$-$\omega$ mixing. The BW parameterization, supplemented with a dipole form factor, for the $\rho$ and $\omega$ mesons was applied again in this analysis, and $R_X$ was extracted to be $0.25\pm 0.01$ for a running $\rho$ width and $0.30\pm 0.01$ for a constant $\rho$ width in the $\rho$ propagator. 
So far, there is no analysis on the $\Xspi$ decay treating the broad $\rho$ resonance and the $\rho$-$\omega$ mixing properly at the same time.

In view of the above discussion, here we perform an analysis of the LHCb data 
for the decay $X(3872)~\to~J/\psi \pi^+\pi^-$,  
where 
a dispersive approach is applied to describe the 
universal nature of the $\pi\pi$ final state interaction (FSI), through which the $\rho^0$ resonance enters. This approach allows us to analyze 
the LHCb data accurately and extract the value of $R_X$ in a reliable manner.
The value found for this important quantity in this way is significantly smaller than those quoted above.

This paper is structured as follows. In Section \ref{Sec:Amp_2pi}, 
we discuss the $\pi\pi$ FSI 
and how it is included in the 
$\Xspi$ amplitude, along with the proper isospin-breaking effects.
Our results of the
 fit to the LHCb data
 are discussed in Section~\ref{Sec:Res}.  
Section~\ref{Sec:3pi} presents our prediction for the isospin-conserving 
$\Xlpi$ decay and the updates on the predictions of the $W_{c1}$ states, the isovector partner of the $\X$. Finally, Section~\ref{Sec:conc} presents a brief summary.

\section{The \texorpdfstring{$\bm{\Xspi}$}{X to Jpsi pi pi} amplitude}
\label{Sec:Amp_2pi}

In this section, we discuss the construction of the decay amplitude used in the evaluation of 
the $\pi^+\pi^-$  invariant mass distribution in the $\Xspi$ decay. We begin with the 
implementation of the $\pi\pi$ FSI. Next, we  discuss the inclusion of 
the factor that encodes isospin breaking and its 
correspondence with the ratio $R_X$, which is the quantity in the focus of this investigation.

\subsection{\texorpdfstring{Universal $\bm{\pi\pi}$}{pipi} FSI}
\label{Subsec:omnes}

The $\pi^+\pi^-$ FSI plays 
an important role in describing the process $\Xspi$. 
In this particular case, the pions 
interact in the $P$-wave ($\ell = 1$). For a given  partial wave, the phase of the $\pi\pi$ FSI amplitude (or pion form factor) in the elastic regime equals to the $\pi\pi$ scattering phase shift $\delta_{\ell}(s)$ modulo $n\pi$ with $n$ an integer (Watson's theorem~\cite{Watson:1954uc}), with $\sqrt{s}\equiv m_{\pi^+\pi^-}$ the invariant mass of the $\pi^+\pi^-$ pair in their center-of-mass (c.m.) frame. 
Consequently, the $\pi\pi$
FSI is described by a universal function called the Omnès function $\Om$~\cite{Omnes:1958hv}, which, in our case, is given in 
terms of the $P$-wave elastic phase shift $\delta^1_{1}(s)$ as
\begin{equation}
\Om = \exp \left[\frac{s}{\pi} \int_{4M_{\pi}^2}^{\infty} ds^{\prime}\,
\frac{\delta_1^1\left(s^{\prime}\right)}{s^{\prime}\,
\left(s^{\prime}-s-i \varepsilon\right)}\right]\, .
\end{equation}
Since we are interested 
in analyzing LHCb data for $\Xspi$, where the $\pi\pi$ invariant mass is limited by the phase space to be $\sqrt{s}\lesssim 0.775$~GeV, inelastic effects can be safely neglected, and we can use the $P$-wave elastic scattering phase shift from Ref.~\cite{Garcia-Martin:2011iqs}. For a treatment of the pion vector form factor including the high-energy region where inelasticities become important (particularly above 1~GeV), we refer to Ref.~\cite{Hanhart:2012wi}. 

In terms of the Omnès function, $\Om$, the amplitude of $\Xspi$ can be constructed as
\begin{align}
\mathcal{M}_{X \to J/\psi \pi \pi}= \mathcal{N}\,
\varepsilon_{ijk}\,\varepsilon^{i}_{\psi}\,\varepsilon^{j}_{X}\,q^{k}_{\pi}\, P(s)\,\Om \,,
\label{Eq:amp2pi}
\end{align}
where $\varepsilon_{\psi}$ and $\varepsilon_{X}$ are the polarization vectors for the $\jp$ and $\X$, respectively, $q_{\pi}$ is the c.m. momentum of the $\pi^+$, and $\mathcal{N}$ represents the overall strength, which will serve as the normalization constant in the fitting later.
The function $P(s)$ appears, since the linear unitarity relation for the form factor
fixes it  only up to a function 
that does not have a right-hand cut,
most easily parameterized by a polynomial.
In Refs.~\cite{Stollenwerk:2011zz,Hanhart:2013vba}, 
the $\pi\pi$ 
FSI was taken into account in the reactions $e^+e^-\to\pi^+\pi^-$ and
$\eta\to \pi^+\pi^-\gamma$ together with
a linear polynomial (see
Ref.~\cite{Gasser:1990bv} for a related discussion). In Ref.~\cite{Kubis:2015sga}, it was demonstrated that
a prominent left-hand cut can call for a second
order polynomial. However, since there
is no obvious meson exchange providing a left-hand cut 
contribution here, we employ
\begin{equation}
P(s) = 1 + \alpha\,s\, .
\label{Eq:pol}
\end{equation}
The slope $\alpha$ will be 
left as a free parameter to be constrained by the fit 
to the LHCb data. 

\subsection{Including the isospin breaking effects}
\label{Subsec:Isobreak}

The Omnès representation discussed above captures only the $\rho$ resonance associated 
with $\pi\pi$ isovector interactions in the elastic region and does not encode any isospin breaking contribution from the $\omega$ meson via $\omega \to \pi^+\pi^-$. The effects of this isospin breaking, typically of $\mathcal{O}(10^{-3})$, are overcome near the 
$\omega$ pole by a factor $M_\omega/\Gamma_\omega\sim 90$ induced by the
$\omega$ propagator (see Ref.~\cite{Daub:2015xja}
for a detailed discussion).
Therefore, it can vary the $\Xspi$ amplitude
significantly, as 
the LHCb data~\cite{LHCb:2022jez} indeed suggest. 

According to Refs.~\cite{Barkov:1985ac,Gardner:1997ie,Leutwyler:2002hm,Hanhart:2012wi,Daub:2015xja}, 
the $\rho$-$\omega$ mixing can be introduced as 
\begin{align}
\mathcal{M}_{X \to J/\psi \pi \pi}= \mathcal{N}\,
\varepsilon_{ijk}\,\varepsilon^{i}_{\psi}\,\varepsilon^{j}_{X}\,q^{k}_{\pi}\, P(s)\,\Om\, 
\left[1+\kappa_X\,G_\omega(s)\right],
\label{Eq:amp2piISOV}
\end{align}
where $G_\omega(s)$ is the propagator of $\omega$,
\begin{align}
 G_{\omega}(s) &= \frac{1}{s - M^2_\omega + i\,M_{\omega}\,\Gamma_{\omega}}\,.
    \label{Eq:omegapropagator}
\end{align}
The parameter $\kappa_X$ encodes
the isospin-breaking effects in the present case, 
$M_\omega$ and $\Gamma_\omega$ stand for the $\omega$ 
meson mass and its decay width, respectively. Here we use a constant width for the $\omega$---we checked that the energy dependence of $\Gamma_\omega(\sqrt s)$, whose explicit expression is shown in Appendix~\ref{sec:AppA}, has negligible effects on the final results. %

Crucial for this analysis is the connection between
the parameters $\kappa_X$ and $R_X$. This is done by performing a matching 
between the amplitude in Eq.~\eqref{Eq:amp2piISOV}, Laurent expanded 
around the $\rho$ pole, and the amplitude corresponding to the decay of  $X$ via  $\rho$ including the $\rho$-$\omega$ mixing, 
given by~\cite{Hanhart:2011tn}
\begin{align} 
 \mathcal{M}_{X \rightarrow J / \psi \pi^{+} \pi^{-}}^{\rm BW}
 = & -g_{X J / \psi \rho}\,\, g_{\rho \pi^{+} \pi^{-}}\,
    \epsilon_{i j k}\,\, \epsilon_X^i\, \epsilon_\psi^{* j}
    \, q_\pi^k\, P(s)\,\notag\\
    &\times G_\rho(s)\left(1-\frac{\epsilon_{\rho\omega}}{R_X}\,
 G_\omega(s)\right) \,,
    \label{Eq:amp_rho}
\end{align}
with $g_{X J / \psi \rho}$ the 
$X(3872)$ coupling to the $J / \psi \rho$ mode.
{The parameter $\epsilon_{\rho\omega}$ measures the $\rho\text{-}\omega$ mixing. Using the results in Ref.~\cite{Colangelo:2023rpc}, its value is determined to be $3.35(8)\times 10^{-3}~\mathrm{GeV}^2$; see Appendix~\ref{App:eps} for details. It turns out the uncertainty of $\epsilon_{\rho\omega}$ has negligible effect compared to the statistical error of $R_X$ from the fitting.} 
In Eq.~(\ref{Eq:amp_rho}),
$G_\rho$ is the propagator of $\rho$ in the BW form,
\begin{align}
 G_{\rho}(s) &= \frac{1}{s - M^2_\rho + i\,M_{\rho}\,\Gamma_{\rho}(\sqrt s)}\,,
    \label{Eq:rhopropagator}
\end{align}
with $\Gamma_{\rho}(\sqrt s)$ the energy-dependent width of $\rho$, as detailed in Appendix~\ref{sec:AppA}.

On the other hand, around the $\rho$ pole, the amplitude 
$\mathcal{M}_{X \to J/\psi \pi^+ \pi^-}$ in Eq.~\eqref{Eq:amp2piISOV} 
can be expanded as
\begin{align}
\mathcal{M}_{X \rightarrow
J/\psi \pi^{+} \pi^{-}}=&\,\mathcal{N} \,\epsilon_{ijk}\, \epsilon_X^i\, \epsilon_\psi^{* j}
\,q_\pi^k\,\mathcal R P(s)G_\rho(s)\notag\\
&\times \left[1 + \kappa_X\, G_\omega(s)\right] + \textrm{regular terms}\,,
\end{align}
where $\mathcal R$ is the residue of the Omnès function at the $\rho$ pole. 
Thus, by performing the matching, we obtain
$\mathcal{N}\,\mathcal R=
-g_{XJ/\psi\rho}\,\,g_{\rho \pi^+ \pi^-}$ and especially
\begin{align}
\kappa_X = -\frac{\varepsilon_{\rho\omega}}{R_X}\,.
\label{Eq:mathcing}
\end{align}
Thus, once $\kappa_X$ is obtained from the fit, Eq.~\eqref{Eq:mathcing} directly provides the value for the 
ratio $R_X$.

\section{Fits to the LHC\lowercase{b} data}
\label{Sec:Res}

Once the decay amplitude associated with the process $\Xspi$ 
is defined, we can write the invariant mass distribution of the 
$\pi^+\pi^-$ pair as
\begin{align}
    \frac{d\Gamma_{X\to J/\psi\pi^+\pi^-}}{dm_{\pi^+\pi^-}} 
    &=\frac{{{p}}_{J/\psi}\,
    q_{\pi}}{32\pi^3 M^2_X}\,\frac{1}{3}
    \sum_{\textrm{spin}}\,
    |\mathcal{M}_{X\to J/\psi \pi^{+} \pi^{-}}|^2\, ,\label{Eq:dist2pi}
\end{align}
where {$M_X=3871.69$ MeV is the mass of the $X(3872)$}, ${{p}}_{J/\psi}$ is the momentum of $J/\psi$ in the $X(3872)$ rest frame, $\sum_{\textrm{spin}}$ corresponds to the sum over the polarizations of the
$X(3872)$ and $J/\psi$, and the amplitude $\mathcal{M}_{X\to J/\psi \pi^{+} \pi^{-}}$ 
 is given by Eq.~\eqref{Eq:amp2piISOV}. 

Using Eq.~\eqref{Eq:dist2pi}, { averaged over each bin of $m_{\pi^+\pi^-}$ },we performed a fit to the corresponding $\pi^+\pi^-$ distribution data reported by the LHCb Collaboration~\cite{LHCb:2022jez} to determine the parameters: $\mathcal{N}$, which sets a global normalization constant, $\alpha$, corresponding to the slope of the linear polynomial $P(s)$ in front of the Omn\`es function $\Om$, and $R_X$, which defines the ratio between the $X(3872)$ couplings to the $J/\psi\rho$ and $J/\psi\omega$ channels. 
Moreover, in order to perform the fit, we have considered the experimental energy resolution as well as the efficiency reported in Ref.~\cite{LHCb:2022jez}. { For comparison, we also perform a fit using the BW parameterization in Eq.~\eqref{Eq:amp_rho}.}

\begin{table}[tb]

    \centering
    \caption{Results from the best fit to the LHCb data~\cite{LHCb:2022jez} using the Omn\`es or BW parameterization for the $\rho$ meson.  The uncertainties are propagated from the $1\sigma$ statistical errors of the data. 
    }
    \begin{ruledtabular} 
    \begin{tabular}{l  c c c}
    Parametrization & $\alpha$ (GeV$^{-2}$) & $R_X$& $\chi^2/\mbox{dof}$ \\\hline
     Omn\`es   & {$0.70\pm 0.32$} & {${0.26\pm 0.03}$}& 1.29 \\ \hline
      BW  & {$1.30\pm 0.47$} & {${0.30\pm 0.03}$} & 1.32
    \end{tabular}
    \end{ruledtabular}
    \label{Tab:fitres}
\end{table}

The best fits lead to {$\chi^2/\mbox{dof}=1.29$ for the Omn\`es parameterization and 1.32 for the BW parameterization}, where $\mbox{dof}$ 
denotes the number of degrees of freedom.
The parameter values obtained from the fit, together with the corresponding $1\sigma$ uncertainties propagated from the statistical errors of the data, are listed in Table~\ref{Tab:fitres}. 
We have checked that the parameters, within the uncertainties, are insensitive to the energy-dependence of the $\omega$ decay width, as concluded in Ref.~\cite{Wang:2022vjm}, although in that analysis the energy-dependent case provided a slightly larger value than the constant one. 
The central value of $R_X$ obtained from our fit {using the Omn\`es parameterization} is smaller than the one extracted in Ref.~\cite{LHCb:2022jez}, $0.29\pm0.04$, { which is close to our results using the BW parameterization. 
As the Omn\`es parametrization, which contains not only the $\rho$ pole but also regular terms, is more proper than the BW one, the value of $R_X$ extracted with the Omn\`es parameterization is regarded as our final result.
The visible difference of the central values shows the importance of using a more proper parametrization.
}

\begin{figure}[tb]
	\centering
\includegraphics[width=\linewidth]{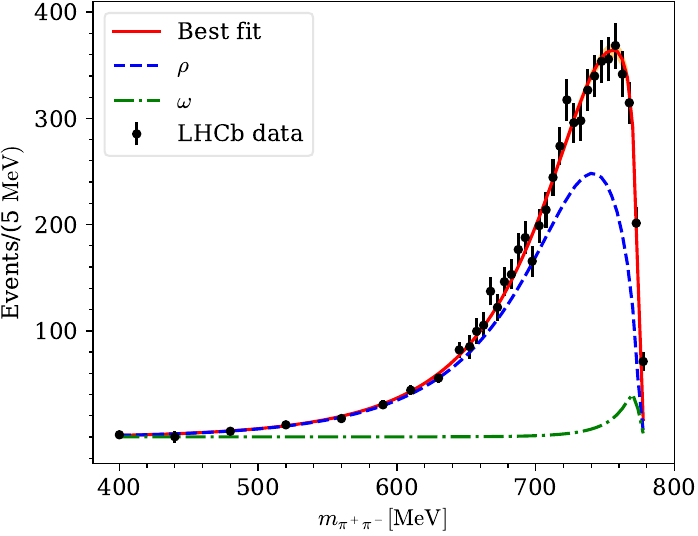}
	\caption{Comparison of the best fit result of the $\pi^+\pi^-$ invariant mass 
    distribution (red solid line), given in Eq.~\eqref{Eq:dist2pi}, with 
    the corresponding data from LHCb Collaboration~\cite{LHCb:2022jez}. The almost invisible band corresponds to the $1\sigma$ error region. 
    The blue dashed line corresponds solely to the $\rho$ meson contribution, 
    while the green dot-dashed one features the $\omega$ contribution 
    to the spectrum, obtained by dropping the unity inside the square brackets in Eq.~\eqref{Eq:amp2piISOV}. Note that due to interference the red distribution is not equal to the sum of the blue and green ones.
 }\label{Fig:dist_2pi}
\end{figure}
In Fig.~\ref{Fig:dist_2pi}, we show the comparison between the lineshape of the 
$\pi^+\pi^-$ distribution in Eq.~\eqref{Eq:dist2pi} (red solid line) 
and the corresponding spectrum measured by LHCb (black dots with error 
bars)~\cite{LHCb:2022jez}. The almost invisible error band is the $1\sigma$ error region corresponding to the uncertainties of the fitted parameters.
An excellent agreement with the data is obtained
across the entire mass range of the spectrum, including the high-energy 
region around the peak at $770$~MeV, which is dominated by the $\rho$ meson. 
This behavior becomes more evident when analyzing the line shape of the 
$\pi^+\pi^-$ distribution considering only the $\rho$ contribution 
(blue dashed line), highlighting a peak precisely in the region where the 
$\rho$ should dominate the spectrum. It is important to emphasize that the 
$\rho$ contribution arises naturally in our amplitude, as it is 
fully encoded in the pion-pion rescattering effects 
captured by the Omn\`es function $\Om$. 
Furthermore, the green 
dot-dashed line in Fig.~\ref{Fig:dist_2pi} corresponds to the lineshape solely 
due to the $\omega$ resonance, which, although small compared to the $\rho$ meson one, is still sizeable to the spectrum under study.

\section{Predictions}
\label{Sec:3pi}

\subsection{The \texorpdfstring{$\bm{X\to J/\psi \pi^+\pi^0\pi^-}$}{X to Jpsi pi+ pi0 pi-} spectrum}

Once $R_X$ is extracted from the data, it can be used to predict the line shape 
of the $\threepi$ mass distribution in the decay $\Xlpi$.
In particular, we will follow the discussion in 
Ref.~\cite{Hanhart:2011tn} in defining the amplitude 
$X(3872) \to J/\psi \threepi$. In this case, the amplitude 
can be divided into two contributions: one due to the 
$\omega$ resonance and the other due to the $\rho$ resonance via isospin breaking, where the quantity 
$R_X$ enters. Thus, around the peak 
of the distribution, which is also close to the $\omega$ 
pole, we have
\begin{align}
    \mathcal{M}_{X \to J/\psi\omega} = & g_{X J/\psi\omega}\, 
    \epsilon_{i j k}\, \epsilon^i_X\, \epsilon^{* j}_\psi\, 
    \epsilon^{* k}_\omega\,\left(1-\epsilon_{\rho\omega}\, R_X\, G_\rho\right)\, ,
    \label{Eq:amp3pi}
\end{align}
where $g_{X J/\psi\omega}$ represents the coupling of $X(3872)$ to $J/\psi\omega$.
The differential decay width corresponding to the $\Xlpi$ decay via the $\omega$ intermediate state reads 
\begin{align}
    \frac{\mathrm{d} \Gamma_{X \rightarrow J / \psi 3 \pi}}{\mathrm{d} m_{3 \pi}}=&\,\frac{1}{4 \pi^{2} M_{X}^{2}} \frac13\sum_{\rm spin}\left|\mathcal{M}_{X\rightarrow J / \psi \omega}(m_{3\pi}^2)\right|^{2}  p_{J/\psi}\notag\\
    &\times\left|G_{\omega}\left(m_{3 \pi}^2\right)\right|^{2} m_{3 \pi}^2 \Gamma_{\omega \rightarrow 3 \pi}\left(m_{3 \pi}\right),
    \label{eq:Gamma3pi} 
\end{align}
with $\mathcal{M}_{X\to J/\psi\omega}$ the amplitude given by 
Eq.~\eqref{Eq:amp3pi} and $\Gamma_{\omega\to 3\pi}$ defined in 
Eq.~\eqref{Eq:omega3pi}.

\begin{figure}[tb]
	\centering
\includegraphics[width=\linewidth]{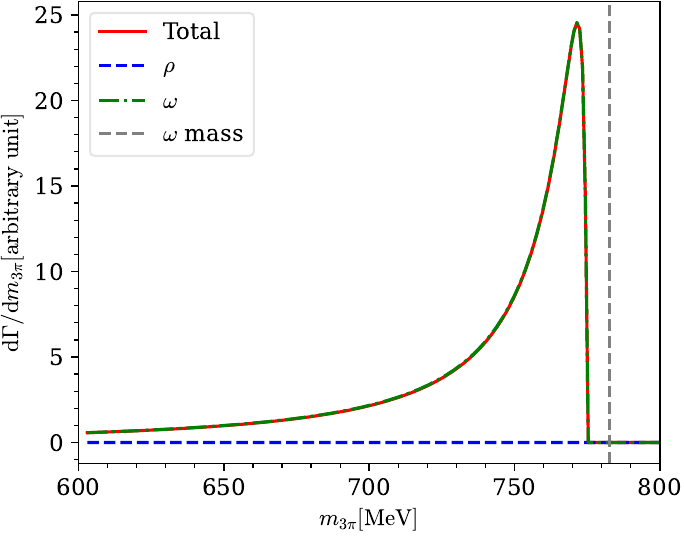}
	\caption{Prediction for the $\pi^+\pi^0\pi^-$ invariant mass distribution of the $X(3872)$ decay, as given by Eq.~\eqref{eq:Gamma3pi}. The $1\sigma$ error band from the errors of the parameters is too narrow to be seen. The gray 
dashed one locates the nominal $\omega$ mass.
 }
\label{Fig:dist_3pi}
\end{figure}

Figure~\ref{Fig:dist_3pi} shows our prediction for the 
$\pi^+\pi^0\pi^-$ spectrum 
from the four-body $\Xlpi$ decay. As can be seen, it exhibits a 
sharp peak in the high-energy part of the distribution, which 
then abruptly drops off due to the phase-space boundary. In this 
region, the distribution is supported only by a small portion of 
the $\omega$ pole (the vertical gray dashed line shows the nominal $\omega$ mass), specifically from 
its tail~\cite{Suzuki:2005ha}, since the $\omega$ nominal mass lies outside the physical boundary allowed by the phase space. 
In addition, unlike the previous case, the contribution from 
the $\rho$ meson is very small and does not affect the line shape 
of the three-pion spectrum, which is not surprising since the $\rho$ contribution to the $X(3872)\to J/\psi\threepi$ is doubly suppressed by the isospin-violating coupling of $X(3872)$ to $J/\psi\rho$ and the small $\rho$-$\omega$ mixing.

It should be noted that in principle
also the isovector state $W_{c1}$, proposed to 
exist in Ref.~\cite{Zhang:2024fxy}, should contribute
to this spectrum, as well as the $\pi\pi J/\psi$ spectrum discussed before as will be exploited in the
next paragraph. Since this state decays predominantly
into $\rho J/\psi$, it could lead to a modification of
the 3$\pi J/\psi$ spectrum via a mixing from the 
$\rho$ to the $\omega$, driven by the same amplitude
already discussed above.
Unfortunately we are not able to generally
quantify this impact here, since the production strength of the $W_{c1}^0$ relative to that of the $\X$ is reaction dependent.

It is clear that the peak in Fig.~\ref{Fig:dist_3pi} is much narrower than the one in the BaBar data~\cite{BaBar:2010wfc}, as observed previously in Ref.~\cite{Hanhart:2011tn}.
Recently, the BESIII Collaboration reported the $\pi^+\pi^0\pi^-$ distribution from the $e^+e^-\to \gamma\pi^+\pi^0\pi^- J/\psi$ reaction~\cite{BESIII:2019qvy}. 
A narrow peak is clearly visible around 0.78~GeV of the 
$\pi^+\pi^0\pi^-$ spectrum, which is mainly due to the $\omega$ meson. 
However, the peak contains not only the events from $\Xlpi$ but also other contributions, such as the $X(3915)\to \jp\omega$, and thus a direct comparison of our prediction with the BESIII data is currently not possible.

\subsection{Updating predictions on the isovector \texorpdfstring{$\bm{W_{c1}}$}{Wc1}}

It was predicted in Ref.~\cite{Zhang:2024fxy} that there should be isovector $D\bar D^*$ hadronic molecules $W_{c1}^0$ and $W_{c1}^\pm$. The quantum numbers of the neutral member is $J^{PC}=1^{++}$.
The prediction has been backed by recent lattice calculations in Ref.~\cite{Sadl:2024dbd}.

The inputs of the calculations in Ref.~\cite{Zhang:2024fxy} is the $\X$ mass and the value of $R_X$ reported by LHCb~\cite{LHCb:2022jez}.
With the new $R_X$ value in Table~\ref{Tab:fitres}, we update the predictions here (for details of the calculations, we refer to Ref.~\cite{Zhang:2024fxy}).
All the poles are located on the unphysical Riemann sheets (RSs) of the corresponding scattering $T$ matrix. 
We use the signs of the imaginary part of the c.m. three-momenta to denote the RSs.
The $W_{c1}^0$ pole is located on RS$_{+-}$ (i.e., the fourth RS) of the $C=+$ $D^0\bar{D}^{\ast 0}$--$D^+D^{\ast-}$ coupled-channel $T$ matrix, and the $W_{c1}^{-}$ pole is located on RS$_-$ (i.e., the second RS) of the $G=+$ $D^0D^{\ast -}$ single-channel $T$ matrix. The pole positions are
{
\begin{align}
    {W_{c1}^0:} &\quad  {3881.7^{+1.0}_{-0.7}+ i (1.2^{+0.8}_{-0.7})~{\rm MeV},}\nonumber \\
    W_{c1}^\pm: &\quad  3862.5^{+6.4}_{-10.3}- i (0.07\pm 0.00)~{\rm MeV},
\end{align}}
where we have only shown the $W_{c1}^0$ pole on upper half energy plane, which is closer to the physical region than the one in the lower half plane~\cite{Zhang:2024qkg}.

The $W_{c1}^0$ pole is {$(10.0^{+1.0}_{-0.7})$~MeV} above the $D^0\bar D^{*0}$ threshold and {$(1.8^{+1.0}_{-0.7})$~MeV}
above the $D^+D^{*-}$ threshold. 
The $W_{c1}^{-}$ pole is {$13.3_{-6.4}^{+10.3}$~MeV} below the $D^0D^{*-}$ threshold.
It is compatible with the lattice QCD result $6.7^{+19.5}_{-\phantom{1}6.7}$~MeV obtained with a pion mass about 280~MeV in Ref.~\cite{Sadl:2024dbd}. 
There is also a shadow pole~\cite{Eden:1963zz,Zhang:2024qkg} of the $X(3872)$ at {$3861.2^{+6.2}_{-10.1}- i( 0.17^{+0.02}_{-0.03})~{\rm MeV}$} on RS$_{--}$ (i.e., the third RS) in the $D^0\bar{D}^{\ast 0}$--$D^+D^{\ast-}$ coupled-channel $T$ matrix.

\section{Summary} 
\label{Sec:conc}

Using dispersion theory to implement the $\pi^+\pi^-$ FSI in the decay $\Xspi$, {we performed an analysis of recent data from the LHCb Collaboration to reinvestigate the isospin breaking effects in this reaction and extracted the ratio between the couplings of $X(3872)$ to the  
$J/\psi \rho$ and $J/\psi \omega$ channels, encoded in the 
parameter $R_X$. The parameter}  provides a measure of isospin violation 
at the $X(3872) \to J/\psi V$ vertex ($V = \rho, \omega$). 
Our result for $R_X=0.26\pm0.03$ is valuable to determine the LECs of the $D\bar{D}^{\ast}$ interaction. 
With the extracted $R_X$ value, we updated the predictions on the isovector $J^{PC}=1^{++}$ $W_{c1}$ poles.

Additionally, we made predictions for the $3\pi$ invariant mass distribution  in 
the four-body decay $\Xlpi$. Measurements of this observable 
are accessible in experiments such as BESIII. Note that there should also be a contribution from the decay of the predicted $W_{c1}^0$
to the spectra discussed in this work. However,
a quantitative prediction for this effect
 needs additional knowledge about the relative  production strength of $X(3872)$ and $W_{c1}^0$ in a given process, which
 could in principle be deduced from an analysis of
 improved data hopefully available in the near future.

\vspace{15pt}

\begin{acknowledgements}
    We would like to thank Jun-Hao Yin and Chang-Zheng Yuan for helpful discussions.
    This work is supported in part by the National Key R\&D Program of China under Grant No. 2023YFA1606703; by the Chinese Academy of Sciences under Grants  No.~XDB34030000 and No.~YSBR-101; by the National Natural Science Foundation of China under Grants No. 12125507, No. 12361141819, and No. 12047503.  U.-G.M. and C. Hanhart, in addition, thank the CAS President's International Fellowship Initiative (PIFI) under Grants No.~2025PD0022 and No.~2025PD0087 for partial financial support.
\end{acknowledgements}

\appendix
{\section{$\rho$-$\omega$ mixing angle}\label{App:eps}

Let $\tilde{\varepsilon}_{\rho\omega}$ represents the mixing in Eq.~(3.3) of Ref.~\cite{Holz:2022hwz}, where the one-photon pole contribution is excluded,
\begin{align}
    F_\pi^{V, e^{+} e^{-}}(s)=\left(1+\tilde\varepsilon_{\rho \omega} \frac{s}{M_\omega^2-s-i M_\omega \Gamma_\omega}\right) F_\pi^V(s),\label{eq:FV}
\end{align}
whose value was determined to be 
\begin{align}
    \tilde{\varepsilon}_{p \omega}= 
    \begin{cases}
        2.00(7) \times 10^{-3} & \text{\cite{Holz:2022hwz}}, \\ 1.99(3) \times 10^{-3} & \text{\cite{Colangelo:2023rpc}}.
    \end{cases}
\end{align}
In the following, we use the most updated value, i.e., the one one in the second line. 
Adding back the one-photon contribution, the complete $\rho$-$\omega$ mixing angle, $\theta_{\rho\omega}$, reads
\begin{align}
    \theta_{\rho\omega}&=\tilde\varepsilon_{\rho\omega}-e^2g_{\gamma\omega}^2=[2.00(7)-0.34(0)]\times 10^{-3}\\
    &=1.66(7)\times 10^{-3},\label{eq:vareps}
\end{align}
where we have used the values of the partial decay width of $\rho/\omega\to e^+e^-$ in Ref.~\cite{ParticleDataGroup:2024cfk} to calculate the couplings of photon and vector mesons,
\begin{align}
    g_{\gamma \rho}&=\sqrt{\frac{3\Gamma_{\rho\to e^+e^-}}{4\pi\alpha^2m_\rho}}=0.201(1),\\
    g_{\gamma \omega}&=\sqrt{\frac{3\Gamma_{\omega\to e^+e^-}}{4\pi\alpha^2m_\omega}}=0.0606(9).
\end{align}

The above $\theta_{\rho\omega}$ in Eq.~\eqref{eq:vareps} is related to the $\epsilon_{\rho\omega}$ parameter in Eq.~\eqref{Eq:amp_rho} as 
\begin{align}
\epsilon_{\rho\omega}=\theta_{\rho\omega}\frac{g_{\gamma\omega}}{g_{\gamma\rho}}{m_\omega^2}=3.35(8)\times10^{-3}\ \rm GeV^2,\label{eq:epsg2}
\end{align}
with a relative error of about $2\%$. 

For comparison, using the formulae in Ref.~\cite{Hanhart:2011tn} and ${\rm Br}{(\omega\to2\pi)}=1.52(8)\%$ extracted in Ref.~\cite{Hanhart:2016pcd}, we get
\begin{align}
    \epsilon_{\rho\omega}\approx\sqrt{m_\omega m_\rho\Gamma_\rho\Gamma_{\omega\to2\pi}}=3.43(10) \times10^{-3}\ \rm GeV^2,\label{eq:epsgam}
\end{align}
with a relative error of about $3\%$.
The two values agree with each other  within $1\sigma$, and the difference in the central values is about $(3.43-3.35)/3.40\approx2\%$.

The uncertainty of $\epsilon_{\rho\omega}$ has little influence compared to the statistical uncertainties of $R_X$ from the fitting, which is about $10\%$, and thus we can safely ignore it.

}

\section{Energy dependence of \texorpdfstring{$\bm{\Gamma_\omega}$}{omega width}}\label{sec:AppA}

For the $\omega$ decay width, we consider two modes, 
$\threepi$, and $\pi\gamma$, with the branching fractions 
$\mathcal{B}[\omega\to 3\pi] = 89.2\%$ and $\mathcal{B}[\omega\to \pi\gamma] = 8.35\%$~\cite{ParticleDataGroup:2024cfk}, 
\begin{align}
    \Gamma_\omega(m)=\Gamma_{\omega \rightarrow 3\pi}(m) 
+ \Gamma_{\omega \rightarrow \pi^0 \gamma}(m).
\end{align}
For the $\pi\gamma$ mode, we have~\cite{Hanhart:2011tn}
\begin{equation}
    \Gamma_{\omega \rightarrow \pi \gamma}(m)=\Gamma_{\omega\to\pi\gamma}^{(0)}\,
    \left[\frac{M_\omega\left(m^2-M_\pi^2\right)}{m\left(M_\omega^2-M_\pi^2\right)}\right]^3\, ,
\end{equation}
with $\Gamma_{\omega\to\pi\gamma}^{(0)} = 0.725$~MeV. For the $\threepi$ mode, we follow Ref.~\cite{Kaymakcalan:1983qq,Kuraev:1995hc} and have
\begin{align}
    \Gamma_{\omega \rightarrow 3\pi}\left(m\right)&=
    \frac{m}{192 \pi^3} \int_{E_+^{\min}(m)}^{E_+^{\max}(m)} d E^{+} 
\int_{E_-^{\min}(m,E_+)}^{E_-^{\max}(m,E_+)}d E^{-}\notag\\
&\quad \mathcal{E}(m,E_+,E_-)\,|F(m,E_+,E_-)|^2,
\label{Eq:omega3pi}
\end{align}
where $E_+$ and $E_-$ correspond to the c.m. energies of the outgoing 
$\pi^+$ and $\pi^-$, respectively, and
\begin{align}
    & \mathcal{E}(m,E_+,E_-)=(E_+^2-M_{\pi^+}^2)(E_-^2-M_{\pi^+}^2) \notag\\
    &-\frac{1}{4}\left[m^2-2m(E_+ +E_-)+2E_+E_-+2M_{\pi^+}^2-M_{\pi^0}^2\right], 
\end{align}
with
\begin{align}
    &E_{+}^{\min}=M_{\pi^+},\quad E_{+}^{\max}(m)=\frac{m^2-M_{\pi^0}
    (2M_{\pi^+}+M_{\pi^0})}{2M},\nonumber\\
    &E_{-}^{\max,\min}(m,E_+)=\frac{1}{2(m^2+M_{\pi^+}^2-2ME_+)} \notag\\
    &\times
    \bigg((m-E_+)(m^2+2M_{\pi}^2-M_{\pi^0}^2-2mE_+)\nonumber\\
    &\pm\Big\{(E_+^2-M_{\pi^+}^2)[m^2+M_{\pi^0}
    (2M_{\pi^+}-M_{\pi^0})-2mE_+]\notag\\
    &\quad\times[m^2-M_{\pi^0}(2M_{\pi^+}+M_{\pi^0})-2mE_+]\Big\}^{1/2}\bigg)\,.
\end{align}
The expression for the amplitude $F(m,E_+,E_-)$ reads 
\begin{align}
    F(m,E_+,E_-)=-\frac{3}{4\pi^2}\frac{g_{\rho\pi^+\pi^-}^3}{F_{\pi}}
    \sum\limits_{a=\pm,0}G_{\rho}(Q_a^2)  ,
\end{align}
with $G_\rho$ the propagator of $\rho$ in the BW form given in Eq.~\eqref{Eq:rhopropagator},\footnote{Since the energy dependence of the width from the $\omega$ meson is tiny, it is safe here to use the BW form for the $\rho$ propagator.}
and
\begin{align}
    Q_\pm^2&=m^2+M_{\pi^+}^2-2mE_\pm, \\ 
    Q_0^2&=M_{\pi^+}^2-m^2+2m(E_++E_-) ,
\end{align}
where $F_{\pi}=92.1$~MeV is the pion decay constant, and the $\rho\pi\pi$ coupling constant $g_{\rho\pi^+\pi^-}$ can be fixed by the experimental
$\rho\to \pi^+\pi^-$ width as ${g_{\rho\pi^+\pi^-}^2}/{4\pi}\simeq 0.50$.
The running decay width of $\rho$ reads~\cite{Kuraev:1995hc}
\begin{equation}
    \Gamma_\rho(m) \simeq \Gamma_{\rho \rightarrow 2\pi}(m) = 
    \Gamma_\rho\left(M_\rho\right) \frac{M_\rho^2}{m^2}
    \left(\frac{m^2-4 M_{\pi}^2}{M_\rho^2-4 M_{\pi}^2}\right)^{{3}/{2}}\,,
    \label{Eq:Gam_run_rho}
\end{equation}
since the $\rho$ decays primarily into $\twopi$ with
$\mathcal{B}[\rho\to\pi^+\pi^-] \simeq 100\%$.

\bibliography{refs}

\end{document}